# Flux dependent MeV self-ion- induced effects on Au nanostructures: Dramatic mass transport and nano-silicide formation

J. Ghatak[1], M. Umananda Bhatta[1], B. Sundaravel[2], K. G. M. Nair[2], Sz-Chian Liou[3], Cheng-Hsuan Chen[3], Yuh-Lin Wang[4], P. V. Satyam[1]*

[1]*Institute of Physics, Bhubaneswar - 751005, India*
[2]*Materials Science Division, IGCAR, Kalapakam, India*
[3]*Center for Condensed Matter Sciences, National Taiwan University, Taipei, Taiwan*
[4]*Institute of Atomic and Molecular Sciences, Academia Sinica, Taipei, Taiwan*

**Abstract**

We report a direct observation of dramatic mass transport due to 1.5 MeV $Au^{2+}$ ion impact on isolated Au nanostructures of an average size ≈ 7.6 nm and a height ≈ 6.9 nm that are deposited on Si (111) substrate under high flux ($3.2 \times 10^{10}$ to $6.3 \times 10^{12}$ ions $cm^{-2}$ $s^{-1}$) conditions. The mass transport from nanostructures found to extend up to a distance of about 60 nm into the substrate, much beyond their size. This forward mass transport is compared with the recoil implantation profiles using SRIM simulation. The observed anomalies with theory and simulations are discussed. At a given energy, the incident flux plays a major role in mass transport and its re-distribution. The mass transport is explained on the basis of thermal effects and creation of rapid diffusion paths at nano-scale regime during the course of ion − irradiation. The unusual mass transport is found to be associated with the formation of gold silicide nanoalloys at sub-surfaces. The complexity of the ion − nanostructure interaction process has been discussed with a direct observation of melting (in the form of spherical fragments on the surface) phenomena. The transmission electron microscopy, scanning transmission electron microscopy and Rutherford backscattering spectroscopy methods have been used.

## I. INTRODUCTION:

When an energetic ion beam impinges on a solid target, many phenomena occur, such as sputtering, recoil atom distribution, defect formation, crater formation on surfaces, etc. The present understanding is basically for an ion impinging a *continuous media (bulk or continuous thin films)* involving incorporation of some kind of spike model (such as a displacement spike leading to thermal spike or coulomb explosion based on the energy regime) [1]. It is known that the beam-target collisions cause energetic recoils, which in turn generate a cascade of secondary knock-on atoms. The elastic collisions induced by the ion bombardment cause ballistic cascade mixing and recoil implantation (RI), wherein target atoms are moved downstream by collisions. The ion-beam mixing (IBM), a process driven by collisions is also possible during the ion-solid interactions. The RI and IBM involve the target – atom redistribution. The other major mechanisms of a target – atom redistribution are, radiation – induced segregation, radiation – enhanced diffusion, and many – body effects (cascade spike effects) [2, 3]. Supporting these experimental findings on ion-solid interactions, several reports are available using molecular dynamics (MD) simulations [4, 5].

Recently, interesting differences for *ion – uniform film* and *ion – nanostructure* interaction phenomena, particularly relating to ion beam mixing, sputtering and surface morphological effects have been reported [6, 7]. Upon irradiation with 1.5 MeV $Au^{2+}$ ions on isolated Au nanostructures on silicon substrate, we found (i) higher probability of crater formation, (ii) larger sputtered particle size and its coverage and (iii) enhanced sputtering yield compared to the continuous films of Au on Si substrate. It was found out that under the low – flux ($\approx 1.3 \times 10^{11}$ ions $cm^{-2}$ $s^{-1}$) irradiation conditions, ion beam induced mixing is found to be absent for continuous Au film on silicon compared to isolated nanostructure system [6]. For this case, mixing at low flux was observed only in case of grazing incidence irradiation geometry (impact angle of 60º) but *no large* mass transport was observed [6].

The ion irradiation being an athermal process, properties of nanomaterials could be tailored, which are otherwise difficult or not feasible by conventional methods [8]. For example, van Dillen *et al.* have shown MeV Au ion irradiation induced anisotropic plastic deformation turning spherical colloids into ellipsoidal shape [9] and the ion-beam irradiation of Pt nanoparticles supported on a $SiO_2$ substrate leads to the burrowing of the particles into the substrate and the result is explained by capillary driving forces and an ion-induced viscous flow of amorphous $SiO_2$ [10]. Earlier, we have also shown the formation of nano-alloys at surfaces, sub-surfaces and in bulk for Au/Si, Au-Ge systems [6, 11]. Burrowing of nanostructures in case of nano–Ag/Si [6, 12] and silicide formation (in the present study) of nano–Au/Si at various fluxes and fluences show the importance of the material properties in ion – nanostructure interactions.

In the present work, a direct observation of mass transport due to MeV self-ion induced effects on Au nanostructures has been reported. In our observations, the effect of the flux (the number of ions falling on the sample of unit area per second) found to have a large impact as no direct mass transport was found for a lower flux values At high flux, a higher rate of defects production would be the main cause for the generation of point defects and creation of rapid diffusion paths. We show below that recoiled and ballistically mixed atoms would be transported due to enhanced diffusion in amorphous silicon system driven by wafer temperature effects. Further, our thickness dependent studies (i.e., mass transport under high flux condition from various sizes of gold nanostructures, from isolated morphology to uniform and continuous morphology) confirmed that the mass transport is not as a consequence of direct cascade induced flow [13].

## II. EXPERIMENTAL METHODS:

Au films of thickness of ≈2.0 nm were deposited by resistive heating method in high vacuum (HV) conditions ($4 \times 10^{-6}$ mbar) on ≈2.0 nm thin native oxide covered Si (111). Irradiations were carried out with 1.5 MeV Au$^{2+}$ ions and at room temperature with different incident ion fluxes ranging from ~ $3.2 \times 10^{10} - 6.3 \times 10^{12}$ ions cm$^{-2}$ s$^{-1}$ over a scanned beam area of 10 mm diameter at two fluence values of $6 \times 10^{13}$ and $1 \times 10^{14}$ ions cm$^{-2}$. The substrates were oriented 5$^o$ off normal to the incident beam to suppress the channeling effects and mounted with silver dug on bulk Cu holder. Transmission electron microscopy (TEM) measurements using the JEOL JEM-2010 facility and the FEI Tecnai G2-F20 scanning transmission electron microscopy (STEM) facility with High Angle Annular Dark Field (HAADF) imaging mode were performed. Cross-sectional TEM (XTEM) samples were prepared by mechanical thinning followed by 3.5 keV Ar ion milling. It is to be noted that the projected range of the 1.5 MeV gold ions (using SRIM) in Au and Si is found out to be ~92 and ~357 nm, respectively [14]. The channeling measurements (RBS/C) were carried out with 2.0 MeV He$^+$ to determine the crystalline quality as a function of depth [15]. The channeling measurements being macroscopic in nature complement the microscopic determination of amorphous nature using the selected area electron diffraction (SAED) and/or with lattice imaging with TEM [16].

### III. RESULTS AND DISCUSSIONS:

We discuss the effect of beam flux (variation from: $3.2 \times 10^{10}$ to $6.3 \times 10^{12}$ ions cm$^{-2}$ s$^{-1}$) on the following aspects of ion – nanostructure interactions:
  A. Dramatic mass transport
  B. Nano-scale mixing leading to alloy formation

*A. Dramatic mass transport:*

Figure 1(a) and (b) depicts the bright field (BF) plan-view TEM and BF XTEM micrographs of as-deposited gold nanostructures for 2 nm thick Au films deposited on Si(111). The effective thickness is determined using RBS measurements. The average size of these isolated nanostructures (using many micrographs shown in Fig. 1(a)) is found to be 7.6 ± 1.5 nm and the average height (using several XTEM BF images as shown in Fig. 1(b)) is found to be 6.9 ± 0.8 nm. The discontinuity (or isolation) of Au nanostructures in the substrate is evident from the above figures. The Figures 1(c) and (d) depicts *direct observation of Au atomic mass transport inside Si* following ion irradiation. Figure 1 (c) shows a conventional XTEM BF image corresponding to the specimen that is irradiated to a fluence of $6 \times 10^{13}$ ions cm$^{-2}$. The irradiation is carried out with 1.5MeV Au$^{2+}$ at a flux of $6.3 \times 10^{12}$ ions cm$^{-2}$ s$^{-1}$. Figure 1(d) depicts the STEM – HAADF image corresponding to Figure 1(c). The confirmation of the presence of Au atoms transport from the nanostructures following the irradiation has been obtained using the STEM – HAADF imaging method. In this method, the inner collection angle of the annular dark field detector is increased beyond the Bragg reflections so that only high – angle scattered electrons contribute to the image [17]. In this mode, the diffraction and phase contrast is significantly suppressed and the scattering cross-section of the electrons distributed in the annular area is roughly proportional to Z$^2$ (where Z is the atomic number) and the bright image contrast indicates the presence of heavy elements directly (Z-contrast). In the Fig. 1(d), the maximum depth to which the Au atoms have been redistributed (for II region of interest) is found to be ≈ 60 nm. The high-resolution lattice image and SAED from the implanted Si show that silicon crystal has already been amorphized (discussed later).

We have considered two depth regimes, the first one being near the surface where brighter contrast was seen (like region I in figure 1(d)) and the other at larger depth where cluster of atoms appeared to be redistributed with lesser contrast (like region II in figure 1(d)). The average push-in depth for the near surface regime is found to be ≈ 9.7 nm with maximum and minimum depths of 13.8 and 5.1 nm, respectively (about 31 regions where the near surface redistribution was found and have been taken into account). The average depth of mass transport is 31.6 nm with maximum and minimum depths of 60.0 and 13.5 nm respectively (about 45 regions have been taken into account). Figure 1(e) and 1(f) show the integrated intensity line profiles representing the Au atoms as a function of depth for the regions I and II as mentioned in the Figure 1(d). Firstly, the area of region of interest is sliced horizontally with a step of 0.3 nm and length 25 nm for region I and length of 44 nm for region II (Figure 1(d)) and the pixel counts are integrated in horizontal direction for each depth step of 0.3 nm. Intensity from the background signal has been subtracted from same area for equal length and depth step. The error bars shown in Fig. 1(e) and (f) are determined by taking the square root of integrated intensity. The anomalous nature is clear from the peaks in the Au atom distribution profile. It is evident that the mass has been transported into the substrate up to a maximum depth of ≈ 17 and 60 nm for the regions I and II respectively. Figure 1(e) and (f) clearly show the non-uniform and anisotropic distribution of representation of Au atoms along the depth of Si.

To understand the mass transport, we have compared the experimental profiles shown in figure 1(e) and (f) with SRIM recoil profiles. Figure 2(a) and (b) shows the depth profiles of recoiled Au atoms generated using Monte Carlo simulations using the SRIM package with 1.5 MeV Au ion bombardments on 2.0 and 20.0 nm thick Au film on Si substrate respectively [14]. Schematic of simulation conditions are shown in the inset. In total, 99999 incident ions have been used to generate these profiles. For both cases, the maximum recoil depth for the Au atoms into Si arising from the top film was found to be ≈ 3.5 nm from the interface and is independent of top Au film thickness. With the increase of top Au film thickness, only the number of Au atoms recoiled per incident ion (i.e. the fraction of Au atoms) increases. Using the SRIM simulations, the energy required for recoiled events to reach a depth of 60 nm into Si would be about ≈160 keV [14]. The energy losses obtained from the SRIM simulations can be used to estimate the required energy for Au atoms to reach a depth of 60 nm. The nuclear energy loss and the electronic energy loss for 1.5 MeV Au ions in Au are 9.5 keV/nm and 2.5 keV/nm, respectively [14]. From the XTEM micrograph, the average island thickness (normal to the surface) is ≈ 7.0 nm (Fig. 1(b)). Total energy loss for 1.5 MeV Au ions in 7.0 nm thick gold targets is about 84 keV (taking total energy loss as the sum of nuclear and electronic energy losses). Assuming that this is the maximum energy available for forward Au atoms (from the nanostructure), it is *not* possible to explain the depth that is achieved by recoiled Au atoms. Statistically, a head-on collision would have much more energy transfer which is possible with a very less probability and hence these are not considered. Also, the theoretical calculations of the mean energy of atoms in a cascade show that most recoils are produced near the minimum energy necessary to displace atoms, $E_d$. Due to the low-energy stochastic nature of these displacement events, the initial momentum of the incident particle is soon lost, and the overall movement of the atoms in a collision cascade becomes *isotropic* [2]. Our observation of mass transport is *anisotropic* in nature. Hence, the observed mass transport would *not be just* the recoiled events. It should be noted that SRIM simulations are done at 0 K (no thermal effects are taken into account) and take only the binary collision into account and hence no non-linear cascade events are taken into consideration. In simple terms, one may consider a reduced $E_d$ value for taking non-linearity into account and this could enhance the number of recoiled events.

The mechanism to understand the dramatic mass transport could be the following: (i) collision cascades drive gold atoms from the film into the substrate through ballistic mixing and recoiled implantation processes. (ii) This is followed by amorphization in the substrate that occurs at lower fluence at high flux condition. (iii) The effective wafer temperature due to incident beam power, drives the mass transport (enhanced diffusion occurring for gold atoms in *amorphous*-silicon system). Our thickness dependent studies confirmed that the mass transport is not as a consequence of direct cascade induced flow [13]. It would be interesting to study the mass transport using the molecular dynamics simulations. More details on the role of high flux, wafer temperature, nano-silicide formation are described below.

*(i) Role of high flux:*

The experimental evidence of flux dependency on the mass transport, mixing and lateral diffusion is shown in Figure 3. Fig. 3(a), 3(b) and 3(c) are the XTEM micrograph of the 2 nm Au film after irradiation at a fluence of $1\times 10^{14}$ ions cm$^{-2}$ at a flux of $3.2\times10^{10}$, $3.2\times10^{12}$ and $4.7\times10^{12}$ ions cm$^{-2}$ s$^{-1}$ (corresponding current densities are 0.01, 1.0 and 1.5 µA cm$^{-2}$) respectively. Fig. 3(d) represents the high-resolution TEM (HRTEM) image taken from the same sample as shown in Fig. 3(c) at the interface in one of the ion-beam-mixed region. It is evident that at low flux up to $3.2\times10^{12}$ ions cm$^{-2}$ s$^{-1}$, no mass transport of Au atoms at the fluence up to $1\times10^{14}$ ions cm$^{-2}$ at normal incidence was observed. SAED pattern in the inset of fig 3(b) indicates that up to this flux (i.e., $3.2\times10^{12}$ ions cm$^{-2}$ s$^{-1}$), the substrate silicon is still retains its crystalline nature. The agglomeration of Au clusters on the surface is seen in Fig. 3(a) and (b) as the size of Au nano islands has been increased up to a maximum value of 25.0 nm, which is bigger compare to the as-deposited case. It has already been shown that at same fluence and at flux of $1.3\times10^{11}$ ions cm$^{-2}$ s$^{-1}$ (low flux conditions), only agglomeration of Au clusters on the silicon surface was observed [6]. But in Fig. 3(c), the similar effects are more prominently shown which has been shown in Fig. 1(c) and 1(d). In this case, the amorphization of Si has also been observed, which is evident from the SAED pattern (see inset of fig 3(c)). From HRTEM images (e.g. Fig. 3(d)), lattice spacing was found to be $0.303 \pm 0.005$ nm, which is closely matching with (120) plane of hexagonal phase of Au-Si alloy ($Au_5Si_2$) [18]. Even though this value is closer to silicon lattice spacing of 0.313 nm, but the amorphous nature from SAED pattern shown in inset of fig 3(c) confirms that no lattice spacing would be seen from the silicon substrate matrix. From Figure 3, it is evident that mass transport is happening only where Si substrate is found to be amorphous, and also at high flux conditions. Even at fluence of $6\times 10^{13}$ ions cm$^{-2}$ with flux $6.3\times10^{12}$ ions cm$^{-2}$ s$^{-1}$, substrate was found to be amorphous [15]. Figure 4(a) shows the SAED pattern from the circular region shown in Figure 1(c). The diffused rings for amorphous Si (a-Si) are clearly seen in the diffraction pattern. To confirm the macroscopic amorphous nature and the depth to which the amorphous nature exists, RBS/C measurements were carried out. Figure 4 (b) shows the random and aligned spectrum from irradiated nano – Au/Si system. By converting the scattering energy of backscattered He ions into depth, it is found that a region of ≈ 620 nm thick from surface to bulk Si is amorphous in nature [15]. Figure 4(c) shows the lattice image obtained from one such mass-transported area like a rectangular region in Figure 1(c). As mentioned earlier for the case and as shown in Fig. 3(d), a lattice spacing of $0.305 \pm 0.004$ nm is found to be present which is closely matching with the (120) plane of the hexagonal $Au_5Si_2$ structure. Here also the amorphous nature of Si (confirmed in shown in both figures 4(a) and 4(b)) rules out the possibility of being Si lattice spacing. The HRTEM images confirm the formation of crystalline nano-gold silicide at sub-surfaces.

The amorphization at lower fluence with higher flux and its correlation with mass transport is explained as following: The main difference in high flux irradiation is the rate of defects creation, which in turn influences the amorphization process in semiconductors. It is already established that the defect creation increases, particularly in semiconductor, with the increase of ion flux [19]. It was also experimentally verified that at a given temperature, amorphization can be achieved at lower fluence with higher flux compared to that of lower flux [20]. But after certain upper limit of flux, dynamical annealing starts where the annihilation of created defects is faster than creation of defects. At this regime of flux higher fluence is required to achieve amorphization. Experimentally it was found that for a fluence above $\sim1\times10^{12}$ ions cm$^{-2}$ and fluxes $\sim1\times10^{11}$ - $1\times10^{15}$ ions cm$^{-2}$ s$^{-1}$, the influence of dynamical annealing in semiconductors decreases with increasing flux. It is to be noted that the flux used for the present experiment is well within this range and dynamical annealing is insignificant in present experimental condition. This is so as the displacements production (vacancy-interstitial pair) rate (P) is proportional to the incident ion flux, which can be expressed as [21]:

$$P(x) = \frac{0.8}{2NE_d}\left(\frac{dE}{dx}\right)_n \dot{\Phi} \qquad (1)$$

where, N is the atomic density of the substrate, $E_d$ is displacement energy of the substrate, $\dot{\Phi}$ is the ion flux, $\left(\frac{dE}{dx}\right)_n$ is the nuclear energy loss per unit depth. Our previous and present experimental results support this fact. In our previous experiments, the flux was $1.3 \times 10^{11}$ ions cm$^{-2}$ s$^{-1}$, for which the critical fluence for complete amorphization was $\approx 2 \times 10^{14}$ ions cm$^{-2}$ for 1.5 MeV Au$^{2+}$ ions while partial amorphization occurred at a fluence of $6 \times 10^{13}$ ions cm$^{-2}$ [16]. The present experiments have been carried out at a flux $6.3 \times 10^{12}$ ions cm$^{-2}$ s$^{-1}$ where a complete amorphization of Si seen at a fluence $6 \times 10^{13}$ ions cm$^{-2}$. From the modified Kinchin – Pease relation (equation 1), one can estimate the production of displacement rate, which is almost $\approx 50$ times higher for flux of $6.3 \times 10^{12}$ ions cm$^{-2}$ s$^{-1}$ compare to the flux of $1.3 \times 10^{11}$ ions cm$^{-2}$ s$^{-1}$. This higher rate of defects production would be the main cause for the generation of point defects and creation of rapid diffusion paths.

**(ii) Role of wafer temperature:**

Another important factor is the ambient wafer temperature due to high beam flux during irradiation. The ambient wafer temperature according to the prescription of Perry has been calculated with the simplest model where the part of input heat due to beam will be absorbed by the wafer and rest of the part, will be radiated only from front surface [22]. The justification of the assumption has been explained by Nakata and with these assumptions, experimentally Nakata found the value of emissivity of the Si wafer to be, $\varepsilon = 0.3$ [23]. The ambient wafer temperature (T) as a function of time (t) during irradiation due to only beam flux (note that there was no external source of heat to the sample) can be obtained from the following equation:

$$P_b dt - \varepsilon\sigma A(T^4 - T_s^4)dt - A\rho LCdT = 0 \qquad (2)$$

where, the first term is the input energy due to beam heating, the second term is the amount of heat radiated from the front surface of the wafer and the third term is the amount of heat absorbed by the wafer. Notations have their usual meaning as they appeared in ref [23]. The equation (2) can be simplified as:

$$\frac{dt}{dT} = \frac{1}{a(T^4 - b)} \qquad (3)$$

where $a = -\varepsilon\sigma/(\rho LC)$ and $b = T_s^4 + P_b/(\varepsilon\sigma)$. In the present experiment, we have used the beam power densities of 0.015, 0.3, 1.5, 2.25 and 3.0 W cm$^{-2}$ corresponding to the fluxes of $3.2 \times 10^{10}$, $6.3 \times 10^{11}$, $3.2 \times 10^{12}$, $4.7 \times 10^{12}$ and $6.3 \times 10^{12}$ ions cm$^{-2}$ s$^{-1}$, respectively. The thickness of the silicon wafer (L) is 0.05 cm and rests of the parameters are given in ref [23]. In the present work, we have used the maximum beam power density 3.0 W cm$^{-2}$ for the flux $6.3 \times 10^{12}$ ions cm$^{-2}$ s$^{-1}$. For fluence $6 \times 10^{13}$ ions cm$^{-2}$, the temperatures calculated is found to be 1125 K at a flux of $6.3 \times 10^{12}$ ions cm$^{-2}$ s$^{-1}$. But for the same fluence of $6 \times 10^{13}$ ions cm$^{-2}$, the calculated temperatures of wafer would be 650 and 360 K at lower fluxes of $6.3 \times 10^{11}$ and $3.2 \times 10^{10}$ ions cm$^{-2}$ s$^{-1}$, respectively. Accordingly, for the same amount of fluence, the wafer temperature due to beam flux is more in higher flux.

The temperature rise is nothing to do with amorphization (as long as flux is not sufficient to cause dynamical annealing), but very important in the context of diffusion. The diffusion can be categorized into two regimes: temperature dependent (thermal diffusion) and temperature independent (radiation enhanced diffusion). In the temperature dependent region, thermal diffusivity is proportional to the temperature. But there is a critical temperature ($T_c$) beyond which radiation enhanced diffusion (RED) found to be significant. Critical temperature ($T_c$) is related the average cohesive energy ($\Delta H_{coh}$) and can be expressed as $T_c = 100 \Delta H_{coh}$ as demonstrated by Cheng [24]. For Au-Si system, the $\Delta H_{coh} = 4.22$ eV/at [25], which gives $T_c = 422$ K in this system which is less compared to irradiation at high flux and hence RED will be significant at this condition. While at low flux, the wafer temperature is less than $T_c$ and hence RED will be insignificant. It should also be noted that the thermal conductivity at room temperature for Au is higher than that of the Si, which also play a role in providing the transient temperatures for mass transportation.

*(iii) Range estimation using an effective diffusion model:*

The observation of faster amorphization (at lower fluence with higher flux) in Si at higher flux is very important to understand the mass transport of Au atoms in Si. A marker layer model has been considered to get an estimation of the diffusion length of Au atoms due to ballistic cascade mixing [2, 26]:

$$D_{cas} t = 0.067 \frac{F_D(x) \Phi}{N E_d} <r^2> \qquad (4)$$

where $D_{cas}$ is the effective diffusivity of the target atoms for a collision induced random walk process, $F_D(x)$ is the damage energy per unit length at distance, $\Phi$ is fluence and $<r^2>$ is the mean squared range of the displaced target atoms. It is well known fact that upon ion irradiation, target atoms do exhibit recoils, which is an instantaneous process. SRIM simulation shows that maximum recoil depth of Au atoms in Si for 2 nm Au/Si is about 3.5 nm (Fig. 2(a)) while irradiated with 1.5 MeV Au ions at normal incidence. The recoiled Au atoms into Si substrate may be assumed as a marker layer. It is now well-established fact that diffusivity of Au in *a*-Si is six orders more than that of crystalline silicon (*c*-Si) [27]. The underlying mechanism behind this enhanced diffusion is the kick-out mechanism [28] where an interstitial Au atom knocks out a Si atom from its lattice site thereby becoming substitutional. Very recently, Au diffusion in *a*-Si has been studied extensively where it was clearly shown that Au diffusion is more in a-Si than in *c*-Si and any intermediate crystalline layer (that is partially *a*-Si) will suppress the effect [29]. In this case, transient wafer temperature also plays a crucial role as, for the same fluence, the wafer temperature is more for higher flux and hence value of thermal diffusivity also will be higher. In demonstrating the effects of thermal nature, the value of $<r^2>$ is determined by treating the $D_{cas}$ in equivalent terms with $D_{th}$ [2]. With a value of $F_D \approx 1760$ eV/nm, N = 50 atoms/nm$^3$, $E_d \approx 13$ eV (for silicon), the calculated and values of $<r^2>$ using equation (4) are tabulated in table 1. $F_D$ was calculated using SRIM simulations and the average value has been taken over a depth of 140 nm inside Si where displacements per atom (DPA) were found to be greater than unity. The Au diffusivity at particular temperature has been taken from the report of Poate *et al.* [30]. From the table, the calculated values of $<r>$ are in broad agreement with the present experimental results. But in this respect, it is to be noted that the time '$t$' and the $\Phi$ (fluence) have been taken as total irradiation time and total fluence. But in reality, the time of complete amorphization and effective fluence due to which dramatic diffusion has taken place is less than '$t$' and '$\Phi$'. This could be one of the reasons why calculated values are higher than that of experimental values.

*B. Nano-scale mixing: gold silicide formation*

There are two theoretical models exist in the literature, one is ballistic model [26] in which mixing occurs through a process of cascade mixing through the interface of two layers or between incoming ions and the target atoms and the another is thermal spike assisted mixing [31, 32] in which mixing occurs through interaction of spikes to each other. It is experimentally [33, 34] verified that target with average atomic number lower than 20 (low mass and low density), thermal spike mixing will be nearly nonexistent and ballistic mixing process will be dominated. It is now well established that a cascade of zones of high collision density (displacement cascade), which evolve into thermal spikes when the energy is thermalized between all the atoms of the local volume. At high energies, the dimensions of the spike is much lower than the distance between the primary collisions, cascade is then formed of sub-cascades containing the spikes. Sigmund et al. [26] did not take into account the break-up of the collision cascade into spatially separated sub-cascades (local spikes) [34]. It has been shown theoretically that for an ion with a high Z and large mass, the total cascade volume will be small and composed of only few spikes which are very close to each other or already overlapped, the mixing rate increases with increase of the probability of cascade overlap [35]. It should be noted that cascades are there regardless of ion flux, but the higher flux make it possible to have more defects and transports paths at nanoscale regime. It appears to us that, the mass transport is occurring *not* as a result of direct cascade induced effects, instead the temperature, amorphization of substrate, defect paths would contribute for the unusual mass transport.

Previously, the experiments on IBM have been carried out with thick continuous films where the energy of the projectile (typically few hundreds of keV) has been chosen such that the thickness of the film is equal to the projected range of the projectile. And keeping these conditions in mind, theories have been developed to explain the mechanisms of IBM. In those cases, ion beam induced mixing is of either ballistic type or due to overlap of thermal spike created at end of range of the projectile (that means at the interface) where migration of substrate atoms into target film and vice versa yields the mixing which required a very high fluence more than $10^{15}$ ions cm$^{-2}$ and additional annealing steps [18]. It is also to be noted that in the present experiments, we have used gold on silicon and Au$^{2+}$ ion with energy 1.5 MeV, so ballistic mixing from the nanostructures alone may not dominate in this case as the projected range of 1.5 MeV ion Au$^{2+}$ is much higher (≈95 nm in Au) than the thicknesses of the films used in the experiments where the projectile ions pass through the film into the Si substrate. That means the spikes created due to this process for all Au films, are at larger depth (≈ 350 nm which is the projected range of 2 nm Au/Si) inside the Si. As the lifetime of these spikes is about few picoseconds, it is almost impossible for that spike to contribute in IBM. But, there is great possibility to generate sub-spikes due to projectile within the volume of Au islands and there could be island-size limited termination of collision cascades or the sub-cascades at the islands substrate interface leading to atomic jump along the lateral and/or beam direction and may yield IBM at the interface.

In the following, we discuss about the phase formation due to ion beam irradiation. Though the heat mixing for Au$_5$Si$_2$ is positive, because of low eutectic temperature (633 K) of Au-Si and higher diffusivity of Au in *a*-Si [27 – 30, 36], the Au/Si system was found to be liable for ion-induced or even spontaneous mixing, as was confirmed not only by this study but also by quite a number of earlier works [6, 18, 37, 38]. Tsaur and Mayer reported details of non-eutectic Au$_5$Si$_2$ phase formation [18]. Under equilibrium conditions, the Au – Si system is a simple eutectic (≈18 at% Si at 363º C) with negligible solid solubility and no intermediate phases. However, by rapid quenching of Au-Si liquid or vapor near eutectic composition, amorphous alloys can be formed. The amorphous phase is unstable at RT and transforms into a metastable crystalline phase, which then gradually decomposes into an equilibrium Au and Si mixture. The effective thermal temperature is more than the temperature required forming eutectic composition (Au$_4$Si), which is about 633 K and hence the absences of eutectic phase in the present experiment. We also carried out the annealing studies on this system around the above temperatures at 633 K (eutectic temperature for Au-Si system) and at 920 K (temperature required for formation of Au$_5$Si$_2$ layer from Au – Si binary phase - diagram). For both these temperatures, *no* inter-mixing or surface alloying is observed for annealed cases. Figure 5(a) and 5(b) show the low magnification and lattice image (HRTEM image) of a system (of 2 nm Au/ Si) that is annealed at 920 K for 30 minutes. After annealing, the average height of the Au islands has been found to be 15.8 ± 2.7 nm. At these higher temperatures, lateral diffusion of Au atoms may cause the increase of average size of the particle. Both these figures confirm that, thermal treatment *alone* would not induce the nano-scale mixing. The native oxide (SiO$_2$) would act as barrier for any inter-diffusion process. From the present results, the range of mass transport has a broad width, anisotropic and long range. As it is not possible to explain the mass transport and mixing just considering the recoil events, we conclude that thermal effects need also to be taken into account.

*Complex nature of ion – solid interactions:*

The complex nature of the ion – nanostructure interaction is evident from the variety of mass transport aspects observed as shown Fig. 6(a) and Fig. 6(B). Figures 6(a) and 6(b) depicts the STEM – HAADF micrographs at other locations of the same system where the flux is $6.3 \times 10^{12}$ ions cm$^{-2}$ s$^{-1}$ and the fluence is at $6 \times 10^{13}$ ions cm$^{-2}$. Formation of small spherical gold nano particles on the surface has been observed in Fig. 6(a) and (b). These spherical particles are found to be gold nanostructures (*no* mixed phase). The formation of isolated particles on the surface could be explained on the basis of presence of a molten state at initial stages, before the diffusion process dominates. It is well known that transient temperature indeed a possibility at the initial stages of ion solid interaction process. One needs to probably take micro-expulsion effects as well to understand the molten stage due to the very high transient temperature. Formation of such hillocks like spherical particles was seen earlier even at low flux irradiations [39]. Figure 6(c) shows the STEM – HAADF image taken at the projected range of 1.5 MeV Au ions in silicon matrix. The implanted gold atoms can be seen in Figure 6(c). Figure 6(c) confirms the use of STEM – HAADF for determination of high Z atoms at low concentrations.

## VI. CONCLUSIONS

We have reported direct observation of the anisotropic mass transport due to MeV Au ion impacts on isolated gold nanostructures. The mass transport depends on the flux and fluence at a given energy. The role of excessive defects (flux dependence) and amorphous nature of the substrate for higher diffusion of Au in *a – Si* are discussed. At a given fluence, change in ion flux changes the sputtering. Wafer temperature due to high beam flux during irradiation also plays a crucial role in mass transport (radiation enhanced diffusion) and sputtering.

## ACKNOWLEDGEMENTS
We acknowledge the cooperation of S. Amirthapandian and P. K. Kuiri during the experiments. We would also like to thank all the staff of both the beam hall at IOP and IGCAR. Thanks are to L. Rehn for fruitful discussions and Prof. Wang for financial support for the visit of IAMS (for PVS).

**Table Captions:**

**Table 1:** Calculated values of diffusivity and average diffused length from the equation (4). Suffix 'a' denotes the formation temperature of $Au_5Si_2$ (from binary phase diagram) and 'b' denotes the value observed in the present experiment and rest of the values has been back calculated from equation (4) and ref [30]. Suffix 'c' corresponds to wafer temperature calculated from equation (3).

**Table 1:**

| Temperature (K) | $D_{th}$ (nm$^2$/sec) | Time (t) (sec) | Fluence (ions/nm$^2$) | <r> (nm) |
|---|---|---|---|---|
| 920 [a] | 100 [30] | 9 | 0.6 | 91 |
| 855 [30] | 44 | 9 | 0.6 | 60 [b] |
| 1125 [c] | 4000 [30] | 9 | 0.6 | 574 |

**Figures Captions:**

**FIG. 1**. (a) and (b) corresponds to the BF Planar and XTEM images of 2 nm Au film on Si(111). (c) and (d) are the BF XTEM and STEM HAADF images taken from same region of 2 nm Au film on Si(111) after irradiated at fluence $6\times10^{13}$ ions cm$^{-2}$ and flux $6.3\times10^{12}$ ions cm$^{-2}$ s$^{-1}$. Regions I and II in (d) show different types of mass transport. (e) and (f) are the experimental distribution profile of Au atom inside Si taken from region I and II respectively. Y-axis is integrated horizontal line scan values from the surface with each 0.3 nm depth step and a proper background correction and x-axis is target depth in nm.

**FIG 2**: (a) and (b) correspond to the recoil distributions of Au into Si generated using SRIM. Au films on Si with thickness 2 and 20 nm have been used as a target and irradiated with 1.5 MeV Au at normal incidence. 99999 Au ions have been used to generate these profiles.

**FIG 3.** Cross-sectional bright field micrographs (low magnification) for irradiated specimens for 2.0 Au/Si systems. The irradiation was done with 1.5 MeV Au$^{2+}$ ions at a fluence of $1\times10^{14}$ ions cm$^{-2}$ with flux (a) $3.2\times10^{10}$, (b) $3.2\times10^{12}$ and (c) $4.7\times10^{12}$ ions cm$^{-2}$ s$^{-1}$ and (d) is the high resolution cross sectional bright field image of fig (c). SAED patterns in inset of (b) and (c) are showing the crystallinity of corresponding substrate after irradiation. All irradiation experiment has been carried out at an impact angle of $5^0$.

**FIG 4**: (Color online) (a) and (c) are the SAED pattern and high-resolution XTEM micrograph taken from circular and rectangular region of figure 1(c) which is irradiated at fluence $6\times10^{13}$ ions cm$^{-2}$ with flux $6.3\times10^{12}$ ions cm$^{-2}$ s$^{-1}$. (b) corresponds to the RBS/channeling measurement on the same sample before and after irradiation at random and aligned condition. Solid line is the simulated aligned RBS data.

**FIG 5**: (a) is the XTEM micrograph of annealed 2.0 nm Au on Si(111) at temperature 920 K for 30 minutes. (b) is the high resolution image of annealed sample shown in (a).

**Fig 6**: (a) and (b) correspond to the HAADF STEM images at the interface of 2.0 nm Au on Si(111) after irradiated at fluence $6\times10^{13}$ ions cm$^{-2}$ with flux $6.3\times10^{12}$ ions cm$^{-2}$ s$^{-1}$. Two different interface modifications show the local melting of the nano islands during ion irradiation. (c) corresponds to the HAADF STEM images of Au distribution at the end of range which is basically implanted ions.

**J. Ghatak et. al. (Phys. Rev. B)**

**Figure 1**

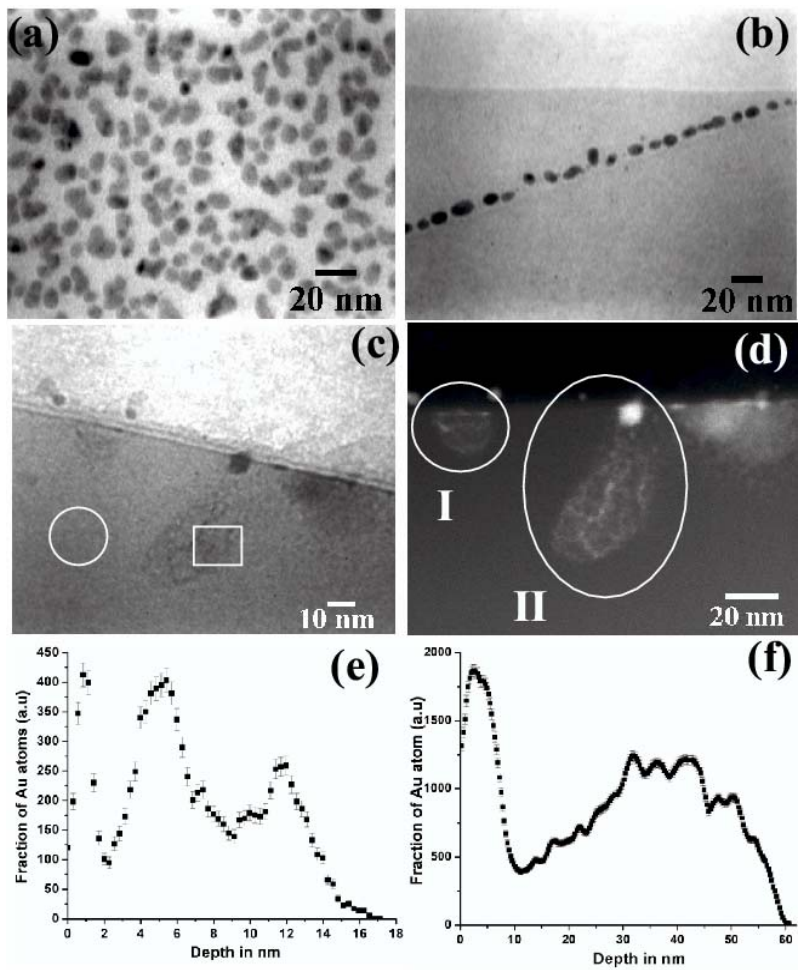

J. Ghatak et. al. (Phys. Rev. B)

Figure 2

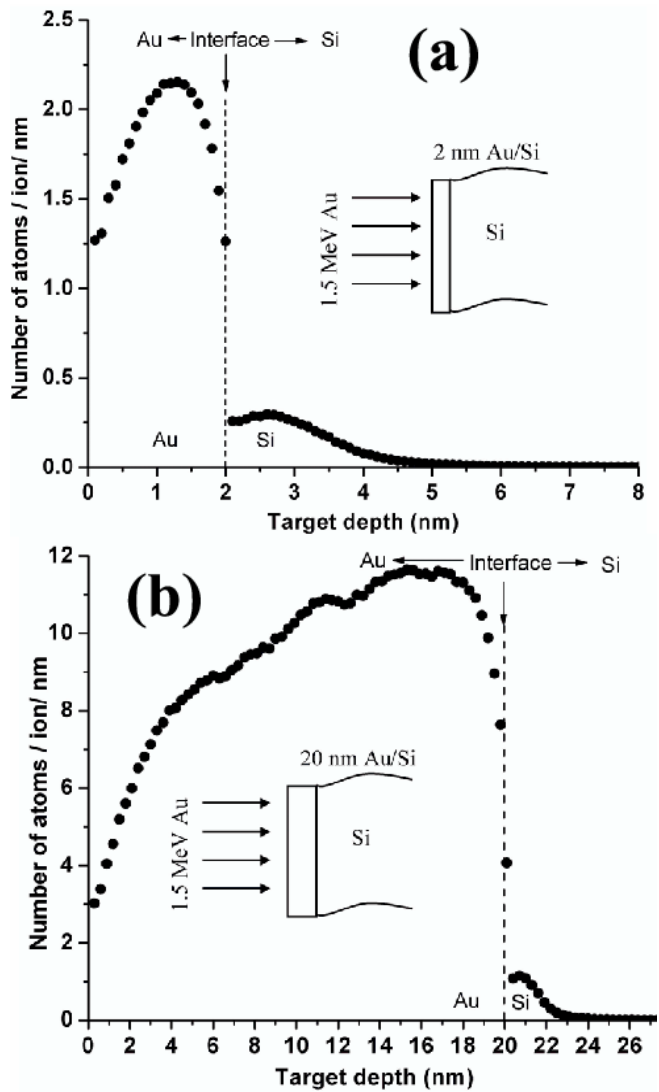

J. Ghatak et. al. (Phys. Rev. B)

**Figure 3**

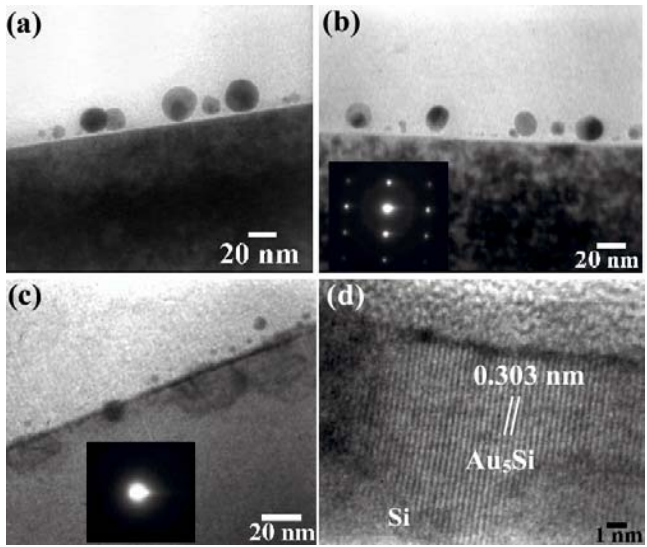

**Figure 4**

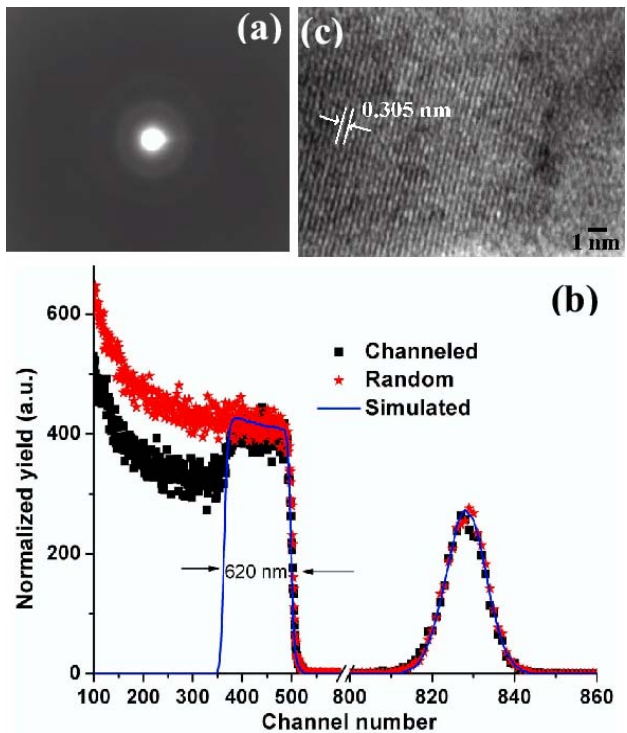

**Figure 5** J. Ghatak et. al. (Phys. Rev. B)

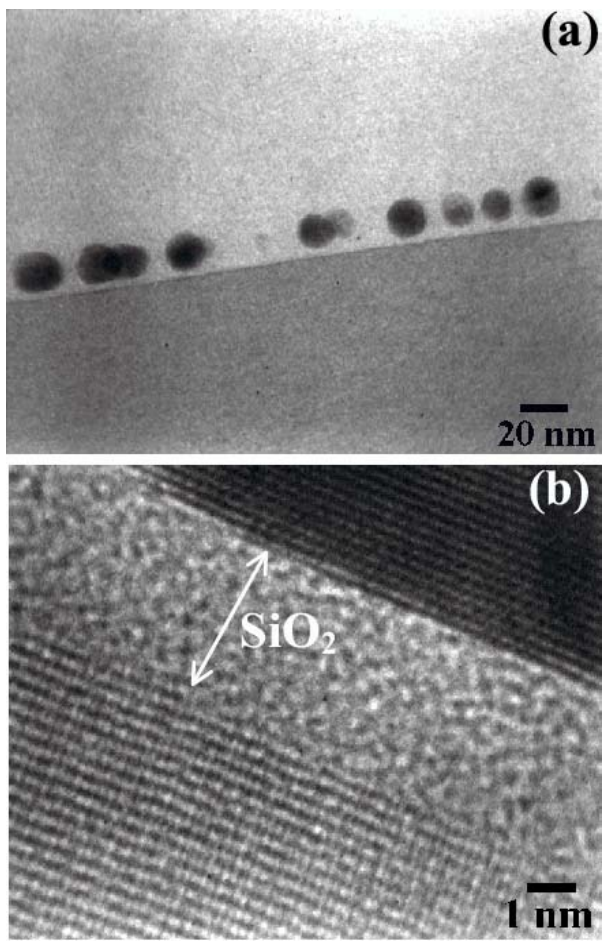

**Figure 6**

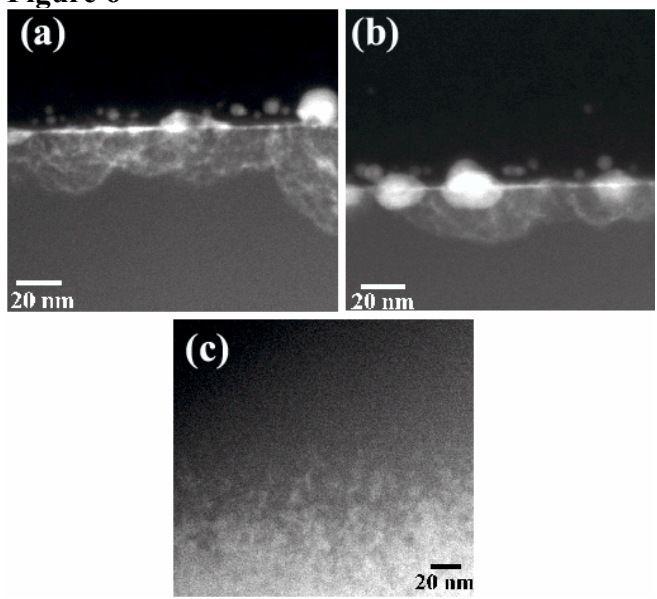